\documentclass[conference]{IEEEtran}
\IEEEoverridecommandlockouts
\usepackage{cite}
\usepackage[linesnumbered,ruled,vlined]{algorithm2e}
\usepackage{setspace}
\usepackage{graphicx}

\SetCommentSty{mycommfont}
\SetKwInput{KwInput}{Input}                
\SetKwInput{KwOutput}{Output}              
\usepackage{booktabs}
\usepackage{caption,subcaption}
\usepackage{textcomp}
\usepackage{xcolor}
\usepackage[breaklinks=true]{hyperref}
\def\BibTeX{{\rm B\kern-.05em{\sc i\kern-.025em b}\kern-.08em
    T\kern-.1667em\lower.7ex\hbox{E}\kern-.125emX}}
\begin{document}

\title{GPU-enabled Function-as-a-Service for Machine Learning Inference\\
{\footnotesize \thanks{This work is partly supported by National Science Foundation awards CNS-1955593 and OAC-2126291.}}
}

\author{\IEEEauthorblockN{Ming Zhao}
\IEEEauthorblockA{\textit{Arizona State University} \\
\textit{mingzhao@asu.edu}}
\and
\IEEEauthorblockN{Kritshekhar Jha}
\IEEEauthorblockA{\textit{Arizona State University} \\
\textit{kjha9@asu.edu}}
\and
\IEEEauthorblockN{Sungho Hong}
\IEEEauthorblockA{\textit{Arizona State University} \\
shong59@asu.edu}
}


\maketitle


\begin{abstract}
Function-as-a-Service (FaaS) is emerging as an important cloud computing service model as it can improve the scalability and usability of a wide range of applications, especially Machine-Learning (ML) inference tasks that require scalable resources and complex software configurations.
These inference tasks heavily rely on GPUs to achieve high performance; however, support for GPUs is currently lacking in the existing FaaS solutions. 
The unique event-triggered and short-lived nature of functions poses new challenges to enabling GPUs on FaaS, which must consider the overhead of transferring data (e.g., ML model parameters and inputs/outputs) between GPU and host memory. 
This paper proposes a novel GPU-enabled FaaS solution that enables ML inference functions to efficiently utilize GPUs to accelerate their computations. 
First, it extends existing FaaS frameworks such as OpenFaaS to support the scheduling and execution of functions across GPUs in a FaaS cluster. Second, it provides caching of ML models in GPU memory to improve the performance of model inference functions and global management of GPU memories to improve cache utilization. Third, it offers co-designed GPU function scheduling and cache management to optimize the performance of ML inference functions. Specifically, the paper proposes locality-aware scheduling, which maximizes the utilization of both GPU memory for cache hits and GPU cores for parallel processing.
A thorough evaluation based on real-world traces and ML models shows that the proposed GPU-enabled FaaS works well for ML inference tasks, and the proposed locality-aware scheduler achieves a speedup of 48x compared to the default, load balancing only schedulers.


\end{abstract}


\section{introduction}

Function-as-a-Service (FaaS) has emerged as a new cloud computing service model which allows users to conveniently deploy and rapidly scale their computing tasks cost effectively. 
However, running machine learning (ML) inference with FaaS functions is limited as the current cloud providers do not support or directly provide FaaS functions to access GPU resources which are critical to accelerate the compute-intensive inference tasks.
For example, AWS Lambda \cite{awslambda} does not provide GPUs to FaaS; Azure functions \cite{azfunction}, FaaS from Microsoft can indirectly access GPUs via GPU-enabled Kubernetes containers, but they cannot share the GPUs.
Therefore, there is an urgent need to enable GPUs on FaaS platforms to allow a wide variety of tasks, including ML inference, to benefit from this service model.

Enabling GPUs on FaaS platforms is imperative to ML inference tasks as they are compute intensive and require low latency to meet the Service Level Agreement (SLA).
ML inference applications in production have stringent latency requirements; for example, providing auto-suggestions in the search bar requires returning the inference results in real-time while users browse for keywords \cite{azfunction}.  
Using GPUs to run ML inference can significantly reduce the latency when input data can be grouped into a large batch and models are designed for parallel computation. 
Taking a batch of input together allows an ML model to take advantage of GPU's parallelism to process them in parallel. 
ML models such as Transformers \cite{wolf2020transformers} translate the sequential computation of recurrent neural networks (RNN) into independent calculations to benefit from GPU parallelization.

Managing the GPU resources in a FaaS platform is challenging as sharing GPUs differs from sharing conventional resources such as CPUs and memory. 
The GPU is designed to maximize a single application's throughput performance by allocating the entire resource to a single GPU process. 
Since the single process has full access to the GPU resource, the GPU expects the application to be programmed to avoid exceeding the available GPU memory and causing out-of-memory (OOM) errors. 
The limited sharing capabilities of GPU are also problematic for the FaaS platform because functions require dynamic sharing of the GPU resources to maximize GPU utilization and function performance. 


The major challenge of GPU-enabled FaaS is to address the above limitations by finding the balance between locality and load-balancing. 
From a locality perspective, GPU-enabled-FaaS can reduce the function latency by serving requests on the same GPU that already has uploaded the model. 
However, favoring locality may increase the average latency of requests because all the requests are forwarded to the GPU that has the model cached while the others are left idle. 
From a load-balancing perspective, GPU-enabled-FaaS can increase GPU utilization by distributing the requests evenly to the GPUs.
However, load-balancing may increase cache misses when handling a workload with a large working set, as models cached on the GPUs cannot get adequately reused and are often evicted out of the limited memory space by the incoming new requests. 

The paper introduces complementary components that allow the existing open-source FaaS platforms to utilize GPU resources and improve the performance of ML inference running as FaaS functions.  
The distributed \textit{GPU Managers} decouple  GPU resource management from the FaaS platform by handling the GPU resources on behalf of FaaS functions; each GPU Manager manages the requests dispatched to a GPU and estimates the GPU's finish time of its queued requests.
The global \textit{Cache Manager} treats the uploaded inference models in each GPU's memory as cache items; it follows the LRU replacement policy to retain the high-locality models in GPU memory. 
The global \textit{Scheduler} uses the estimated finish times and LRU lists from GPU Managers and Cache Manager to schedule and dispatch FaaS functions to GPUs.  

The proposed \textit{locality-aware load-balancing (LALB) scheduler} improves function performance and GPU utilization by balancing the function workload to GPUs and increasing the reuse of cached models when serving the requests. 
Specifically, the scheduler always prioritizes the requests that have their models cached on idle GPUs and can dispatch them out of order in order to promote cache hits.
If the GPU with the cached model is busy, the scheduler uses the estimated finish time of the busy GPU to determine whether the cache hit on the busy GPU has a lower estimated finish time than the cache miss on an idle GPU. 
The scheduler only forwards the cache miss request to an idle GPU when the busy GPUs do not provide a lower finish time with cache hits. 

The performance of the proposed GPU-enabled FaaS is evaluated using real-world traces disclosed by public cloud providers and inference models widely used in production.
The results show that the proposed system is able to allow model inference functions to run on shared GPU resources with a much improved performance.
Specifically, with locality-aware load balancing, the proposed scheduler reduces the average latency and cache miss ratio of the baseline (default load-balancing) scheduler by 80\% and 65\%, respectively. 
With out-of-order dispatch, the scheduler further reduces the average latency and cache miss ratio of the baseline scheduler by 97\% and 81\%, respectively.

The rest of the paper is organized as follows: 
Section~\ref{sec:background} introduces the background and related works; 
Section~\ref{sec:design} describes the major components of the proposed GPU-enabled FaaS system;
Section~\ref{sec:policies} explains the scheduling policies;
Section~\ref{sec:evaluation} presents the performance evaluation; Section~\ref{sec:discussions} discusses additional aspects of the system; and Section~\ref{sec:conclusions} concludes the paper.

\section{background and related works}
\label{sec:background}

\subsection{Function-as-a-Service}

Function-as-a-Service (FaaS) is emerging as a popular cloud computing service model that can provide unprecedented scalability and cost-efficiency to event-driven applications. 
Unlike traditional cloud services that provide VM or container instances for users to deploy and run their entire applications, FaaS provides users platforms for deploying and running the individual functions that compose their applications.
By allowing applications to scale at the function granularity, FaaS allows them to be more scalable---different functions can scale independently and more cost-effective---cloud provider allocates and users pay for only the resources needed by the functions, than traditional cloud services.  


\begin{figure}
\includegraphics[width=\linewidth]{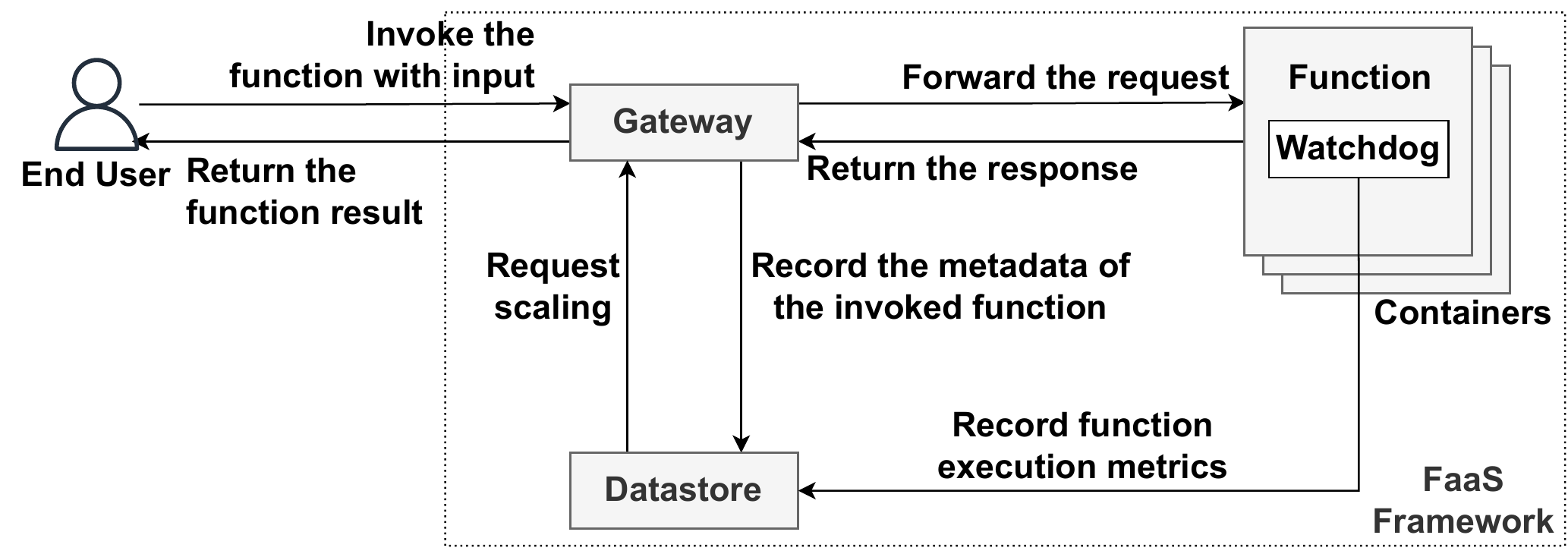}
\caption{General FaaS architecture. It includes three major components: the Gateway provides interfaces to users to deploy and invoke functions, the Watchdog monitors and executes functions, and the Datastore stores function logs and metrics. }
\vspace{-12pt}
\label{fig:faas_default_architecture}
\end{figure}

Figure \ref{fig:faas_default_architecture} illustrates the major components of a typical FaaS framework, including the Gateway, Watchdog, and Datastore, which runs on top of a container orchestration system such as Docker Swarm \cite{docker-swarm} and Kubernetes \cite{kubernetes}.
%
The Gateway is the public route that interacts with the end-users by handling the Create, Read, Update, and Delete (CRUD) operations of functions and invoking the registered functions. 
The Watchdog runs in the background along with the function code on its container to start and monitor the function in the container.
The Watchdog receives the invocation request from the Gateway, executes the function with the given input, returns the response from the function to the Gateway, and stores the status and metrics of the function invocation, such as execution latency to Datastore.
The Datastore stores the log and metrics of the invoked functions. It can also be configured to trigger function scaling actions through the Gateway when the demand for the functions changes dynamically.

An end-user of the FaaS service can write a function without configuring any resources or installing dependencies required to run the function. 
The end-user can use the code template that the specific FaaS platform provides to deploy the FaaS function. 
Once the end-user forwards the function code and the template to the FaaS platform, the platform builds the function by creating a running container that installs the required resources written in the template.
The deployed function is registered as a RESTFUL API, and the end-user can invoke the function by creating an HTTP request or implementing a trigger in other functions or services. 

The benefits of the FaaS paradigm have motivated its use in other computing areas. For example, funX~\cite{funcx} is a function-based scientific computing platform that enables flexible and high performance function execution on federated computing resources; functions have also been explored as the abstraction for computing over heterogeneous edge resources~\cite{jin2020faas}.

Since existing FaaS systems typically do not support GPUs, the scheduling of functions considers their resource requirements mainly in terms of CPUs and memory.
For example, OpenFaaS relies on the underlying container orchestration system such as Kubernetes to queue the functions' containers and schedule them to nodes that can meet their resource needs.

\subsection{GPU Computing}

Graphics Processing Unit (GPU) introduces a different design than conventional processors by offering thousands of simple cores and a high bandwidth memory architecture which can provide massive parallelism to applications such as machine learning.
A GPU has multiple streaming multiprocessors (SM) that contain computing cores, shared cache, and shared memory.
A GPU is an external device that communicates via PCI Express (PCIe) with the host, and data transfer between them is required for the application to utilize the GPU. 



GPUs have three limitations that make it difficult for multiple applications to share the same GPU.
First, the capacity of GPU memory is significantly lower than host memory.
If the processing data size is larger than the available GPU memory capacity, GPU programmers are responsible for managing the active working set in GPU memory.
%
%
Second, although GPU provides fast computations, it incurs extra overhead while transferring data, e.g., model parameters and inputs, from host memory to GPU memory. 
The data transfer overhead of the GPU arises in the PCIe interface as the maximum bandwidth of the current PCIe is much lower (in the order of 100GB/s) compared to the internal memory bandwidth of the GPU (in the order of 1TB/s).

To address the mentioned limitations, it is essential to build a GPU scheduler that can balance requests across GPUs and utilize the models already loaded in GPU memory to service the requests.

\subsection{Deep Learning}

Deep learning applications based on deep neural networks (DNN) are the key solution to many important tasks, such as voice recognition, natural language processing, image classification, and object detection.
The architecture of DNN is represented as the weighted directed graphs where the neurons are grouped as multiple layers, and each connection between neurons communicates with each other.

The main tasks of deep learning comprise training a model and using the model for inference.
The training process updates the weights of each layer iteratively towards a target by running the forward and backward propagation. 
Inference uses the updated weights of each layer to make a prediction based on the input by running the forward propagation.
For image classification, training reduces the difference between the result of the forward propagation and the ground truth label by updating the weights of a model during backpropagation. 
The inference predicts the image class of the image input by translating the input data into a numeric result representing the class.
  
Users typically develop and execute ML tasks using frameworks such as TensorFlow \cite{tensorflow} and PyTorch \cite{pytorch}, which provide convenient high-level APIs and automatically parallelize model training and inference on the available processors, including GPUs.

GPUs offer excellent parallelism for both model training and inference. 
For example, the convolution operation in CNN requires configuring a fixed-sized filter to generate a feature table from the input. 
The filter performs repeated calculations by traversing through the whole image file, and the repeated calculations provide opportunities for GPU to exploit parallelism.
A Larger input batch size also allows more data to be propagated through the neural network, thus allowing more input to be processed in parallel.

Many works have studied GPU-based training of machine learning models. For example, among the recent works, CROSSBOW~\cite{koliousis2019crossbow} is a new single-server multi-GPU system for training deep learning models that enables users to freely choose their preferred batch size; AntMan~\cite{xiao2020antman} co-designs cluster schedulers with deep learning frameworks to schedule model training jobs on large-scale GPU clusters.


This paper focuses on FaaS functions running ML inference because the characteristics of inference tasks are ideal for the FaaS platform to maximize their scalability.
Inference tasks are on demand (often driven by inputs) and relatively short-running (in the order of seconds); training tasks, in contrast, are a lot more time consuming (requiring hours or days) and often not event driven (training data typically does not change frequently).

Advanced model serving systems such as TensorFlow Serving~\cite{tensorflow-serving}, TorchServe~\cite{torchserv}, and NVIDIA Triton~\cite{triton} enable users to streamline the deployment and execution of model inference tasks on individual or cluster of servers. Such a system provides users the interface to deploy models onto the system and get inferences using resources including GPUs. It employs a scheduler or a load balancer to distribute inference tasks over the available resources. 
However, these systems are not based on the FaaS paradigm---FaaS offers unique benefits to model serving, and do not consider GPU cache locality---locality is critical to inference performance.
The goal of this paper is not to develop a new model serving system; instead, our proposed function-based model serving and cache-locality-aware scheduling can be adopted by current and future model serving systems to improve their usability and efficiency.



\subsection{GPU-enabled FaaS}

FaaS functions that use ML inference require significant computation resources and parallelization; therefore, introducing GPUs to FaaS is crucial to improve the performance and scalability of inference tasks.

Kim et al. \cite{8374513} introduced a FaaS platform with GPU support by enabling containers used in the FaaS platform to access GPU directly. %
Although this work shows the benefits of GPU-enabled functions, the functions use the NVIDIA Container Toolkit \cite{nvidiadocker} which restricts multiple containers from sharing a single GPU. 
The function that occupies the GPU may run non-GPU tasks, such as preprocessing the input images while preventing other waiting functions from using the GPU.
Our solution solves the issue of GPU monopolization by enabling the functions to share the GPUs and providing optimized GPU resource management for the functions. 

Naranjo et al. \cite{10.1016/j.jpdc.2020.01.004} addressed the GPU monopolization problem by introducing rCUDA\cite{5547126}, a GPU virtualization framework, to FaaS.
The solution prevents the FaaS functions from directly managing GPUs by intercepting the GPU operations from FaaS functions to the rCUDA interface. 
It decouples FaaS from GPU resource management by relying on rCUDA; consequently, the FaaS platform and rCUDA cannot coordinate to improve performance because they do not share information on pending FaaS requests and GPU utilization.
Our solution, in contrast, builds GPU resource management into the FaaS framework and enables coordinated GPU function scheduling and memory management to improve the function performance and GPU utilization.  

Satzke et al. \cite{satzke2020efficient} implemented features for GPU sharing on top of Knative \cite{knative}, an open-source framework that provides tailored features for managing FaaS functions in Kubernetes.
The solution translates the memory of a single GPU into multiple vGPUs and provides a constraint policy that prevents FaaS functions from oversubscribing the GPU memory. 
However, it focuses on avoiding out-of-memory (OOM) errors caused by GPU memory oversubscription while failing to address the increased overhead caused by multiple functions sharing the same GPU.
Our solution prevents the oversubscription of GPU memory and addresses the resource-sharing overhead with co-designed GPU cache management and function scheduling.

Dakkak et al. \cite{8814494} introduced GPU caching by implementing a daemon process that provisions the GPU memory to functions by intercepting their CUDA requests. 
The solution uses GPU memory virtualization services such as Unified Memory and CUDA IPC to share uploaded models among GPU memory and processes.
The solution prevents OOM in GPU by limiting the memory usage of GPU and using basic cache mechanisms such as LRU to reduce the latency. 
However, the solution does not consider the impact of cache locality on GPU scheduling and always forwards requests to the GPUs with the lowest utilization, which leads to suboptimal performance.   
In comparison, our solution proposes cache-locality-aware scheduling to resolve the conflicts between GPU load balance and cache locality and achieve much improved performance.

Although not specifically designed for FaaS, KubeShare~\cite{yeh2020kubeshare} improves the management of GPUs on a container framework (Kubernetes) by treating GPUs as a first-class resource and allowing them to be scheduled and time-shared by containers;
The AWS virtual GPU device plugin~\cite{aws-vgpu} for Kubernetes allows containers to space-share GPUs using CUDA Multi-Process Service (MPS)~\cite{mps}. 

The issues addressed by the above solutions are complementary to the problem of coordinated GPU memory management and function scheduling that our paper focuses on. On one hand, the increasing shared use of GPUs, as demonstrated by these related works, provides stronger motivation for the effective allocation of the limited GPU memory among concurrent workloads. This is particularly true for model inference tasks which can be significantly affected by the model loading latency. On the other hand, these related solutions can all benefit from our proposed GPU cache-locality-aware function scheduling.
The rest of the paper details the design, implementation, and evaluation results of our solution.

\section{design}
\label{sec:design}

\begin{figure*}
\includegraphics[width=1.0\textwidth]{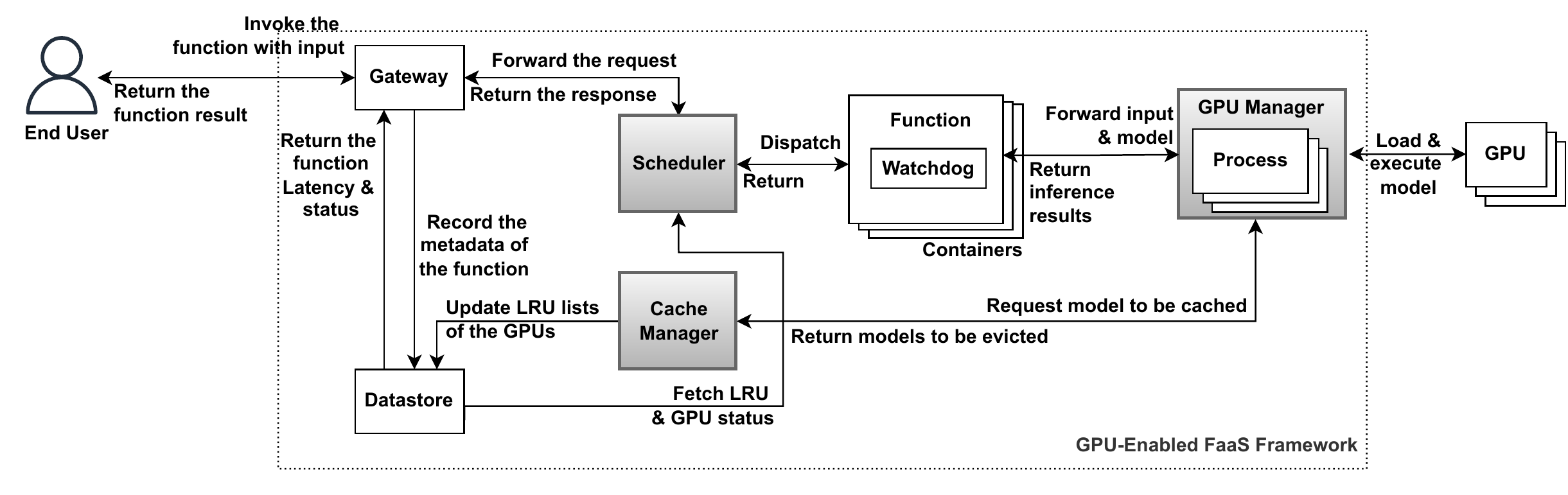}
\vspace{-18pt}
\caption{Architecture of the proposed GPU-enabled FaaS. Unshaded boxes are components from the existing FaaS architecture (with minor changes). Shaded boxes are new components added to the architecture for GPU management, cache management, and locality-aware scheduling.}
\vspace{-6pt}
\label{fig:archicture}
\end{figure*}

\subsection{Architecture}

The proposed GPU-enabled FaaS aims to improve GPU functions' performance, especially for model inference functions, by optimizing GPU scheduling and resource management.
Figure \ref{fig:archicture} shows our framework's complete architecture, which includes three additional components (Scheduler, Cache Manager, GPU Manager), shown as shaded boxes in the figure, which enable and optimize GPU functions upon an existing FaaS framework (shown as unshaded boxes in the figure) such as OpenFaaS.

This GPU-enabled FaaS framework requires minor changes to the Gateway.
The end-user can include a GPU-enable flag in the Dockerfile of the function when registering the function using the Gateway. 
The Gateway checks the GPU-enable flag in the Dockerfile and replaces the interface that the function uses for loading and running a model with a customized interface that redirects those requests to the GPU Manager. 
This change of interface is not visible to the end-user.
%
The interface that needs to be replaced in the function is easy to find as users typically use the common machine learning frameworks for model inference, e.g., torch.load(), model(input) in PyTorch and model.load\_weights(), model.predict(input) in TensorFlow. These APIs are limited, and they are relatively stable across different versions of these frameworks.

\subsection{Scheduler}

The Scheduler decides where (which nodes and which GPUs) to dispatch function requests for the entire FaaS system.
The Scheduler follows a specific scheduling policy that can be enabled when the Scheduler component is first initiated. 
We will discuss the policies in detail in Section~\ref{sec:policies}.
Once the Scheduler decides on a request that needs to be dispatched, it groups the function's information with the GPU address and forwards them to the function's container. 
The function request contains the input data and the registered function's ID that uses the pre-trained model for inference.
The GPU address contains the IP address of the server where the GPU is installed and the device name used to access the GPU on that server.

\begin{figure}[htp]
\centering
\includegraphics[width=\linewidth]{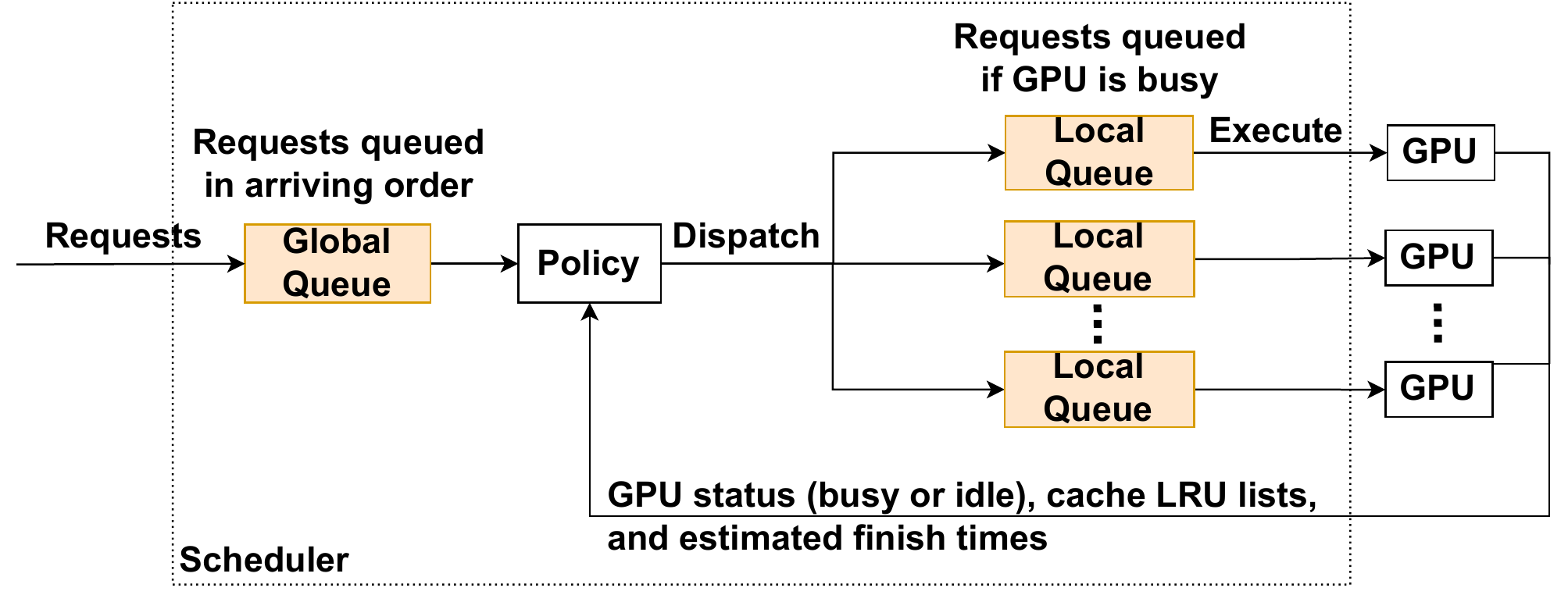}
\vspace{-12pt}
\caption{Architecture of the Scheduler. It maintains a system-wide global queue and per-GPU local queues for scheduling requests according to the given policy and status of the GPUs.}
\vspace{-9pt}
\label{fig:logical}
\end{figure}

Figure \ref{fig:logical} illustrates how the Scheduler dispatches the requests forwarded from the Gateway to GPUs according to the scheduling policy and GPU information collected by the GPU Manager and Cache Manager.
The Scheduler maintains two types of queues: a system-wide global queue and per-GPU local queues.
The global queue is for the entire FaaS system; it stores all the requests forwarded by the Gateway to the Scheduler.
The pending requests in the global queue are sorted by their arrival times. The Scheduler dispatches the pending requests in the global queue to GPUs following the scheduling policy.
There is a local queue for each GPU; it stores all the requests that are scheduled by the Scheduler to the GPU but cannot be immediately executed because the GPU is currently busy.

\subsection{GPU Manager}

GPU Manager runs on each GPU node and manages the GPU processes running on the GPU node. 
GPU Manager runs the PyTorch process on behalf of the function by receiving the inference request from the customized PyTorch API and returning the results to the PyTorch API. 
Each GPU process uploads an inference model when initiating, and after an inference request finishes, it reports the latency to the Datastore. 
When the GPU process is uploading or processing the inference request, GPU Manager reports to the Datastore that the GPU status is busy; and when the GPU process finishes the task, it then updates the status back to idle.
GPU Manager enforces each GPU to run one request at a time and sets the status of the GPU to busy when GPU is processing the request.

GPU Manager communicates with Cache Manager to maintain the models used by the running GPU processes as cache items.  
Given a function invocation request, GPU Manager asks Cache Manager to determine whether the requested model is already cached in the GPU memory (i.e., a cache hit) or not (i.e., a cache miss). If it is a cache miss, it also checks if there are victim models that need to be evicted to make space for the model needed by the new request. 
If there is a victim, GPU Manager kills the process associated with the evicted model.
GPU Manager then starts a GPU process for the new model and uploads and runs the model on the GPU.
If it is a cache hit, there is no need for eviction and the GPU process that uses the requested model is already running.
GPU Manager then forwards the input of the new request to this existing GPU process to run inference.
After the inference finishes, the GPU process returns the result, and GPU Manager returns the result back to the Scheduler.  
The GPU Manager also updates its LRU list of cached models as they are used by its received inference requests.

\subsection{Cache Manager}

Cache Manager runs alongside the Scheduler as a global component and largely follows the LRU replacement policy to manage the models in the memory of each GPU in the system.
The cache items are the models uploaded by the GPU processes to the GPU memory. 
When a function is ready to use GPU, the function requests Cache Manager to return the GPU process that uses the required inference model. 
If the required GPU process exists in GPU Manager, it is a cache hit, as the model used by the GPU process is in the GPU memory and the function can use the existing GPU process and skip the model transfer. 
Otherwise, it is a cache miss and the Cache Manager requests the GPU Manager to create a new GPU process to have the new inference model uploaded to the GPU memory. 

Upon a cache miss, Cache Manager receives a request from the GPU manager that contains the available memory space of the GPU and the ID of the model that is missing in the GPU memory.  
Based on these parameters, Cache Manager then determines the list of victim models that need to be evicted from the GPU in order to make enough space to store the new model, and it chooses the victims according to the GPU's LRU list. 

\subsection{Datastore}

The Datastore stores the estimated latency of each inference request, the LRU list of each GPU, and the status of each GPU.
We use Etcd \cite{etcdsite}, which is already used by the container orchestration system (e.g., Kubernetes), to implement the Datastore.
Etcd is a distributed key-value store that guarantees a high level of consistency applicable in a distributed environment.
The Cache Manager and GPU managers update the aforementioned information in the Datastore, and the Scheduler uses it to decide the optimal GPUs to dispatch the requests.

\section{scheduling policies}
\label{sec:policies}



The default schedulers in a FaaS system mainly consider the load balancing of CPU and memory resources.
In this work, we propose a locality-aware and load-balancing (LALB) scheduler that addresses the unique challenges in scheduling model inference functions on GPUs. 
Besides using the load balancing feature to improve GPU utilization, the scheduler treats the models uploaded to the GPU as cache items to combine the locality-aware feature to enhance the performance of inference functions and the utilization of GPUs.
On one hand, the locality-aware feature in the LALB scheduler reduces the latency of executing a function by forwarding the request to the GPU with the cached model to avoid the overhead of uploading the model. 
On the other hand, the LALB scheduler also considers load balancing and allows cache miss to happen even if a function's model is cached on a busy GPU but would be faster to execute on an idle GPU. 
This design inherently allows popular models to be replicated over multiple GPUs to handle the demand of using these models for inference; it also allows these models to be naturally evicted as the popularity shifts over time in the workload.

\subsection{Locality-aware Load-balancing (LALB)}
\label{sec:lalb}

The locality-aware and load-balance scheduler (LALB) is invoked only when at least one request is waiting in the global queue and at least one GPU is idle.
Algorithms \ref{alg:lalbo3} and \ref{alg:localfunc} explain how the scheduler considers both the GPUs' load balance and the models' locality in the GPU memory. 

First, the LALB scheduler gets the request from the head of the global queue and checks for available idle GPUs that can generate a cache hit, i.e., have the requested model already stored in their memory.
If the request can be a cache hit, the request is dispatched to one of the idle GPUs with the cached item (Algorithm \ref{alg:lalbo3} Line 8). 
After the LALB scheduler finds out that there are no available cache hits on the idle GPUs, it searches for a possible cache hit on the busy GPUs.

If there is a cache hit on a busy GPU, the LALB scheduler compares the estimated finish time of this request for cache hit on this busy GPU to that for cache miss on an idle GPU.
The former includes the time to wait for the busy GPU to finish its current request (and requests already queued in its local queue) and the inference time of the new request.
The latter includes the time to upload the requested model to an idle GPU and perform the inference.
If cache hit on the busy GPU provides a lower estimated finish time than cache miss on an idle GPU, the request is scheduled to the busy GPU and moved to its local queue (Algorithm \ref{alg:localfunc} Line 12). 
When this GPU becomes idle, it always executes the requests already in its local queue before considering any request in the global queue.
On the other hand, when no GPUs with the cached model can produce a lower finish time than an idle GPU, the LALB scheduler dispatches the request to one idle GPU, creating a cache miss (Algorithm \ref{alg:localfunc} Line 17). 

The latencies of uploading the model and running the inference are collected by profiling each unique model on the GPUs in the system.
Table~\ref{tab:model_table} lists the upload time and inference time of the models used in our evaluation.
The upload time depends on only the model size; the inference time depends on the model and the batch size which can be profiled using simple regression methods. 

\begin{algorithm}
\DontPrintSemicolon
\BlankLine
\KwInput{
 \;
 \Indp 
 The list of idle GPUs $idle\_GPUs$ (sorted by frequency) \;  \BlankLine
 The list of busy GPUs $busy\_GPUs$ \;  \BlankLine
 The list of pending requests $requests$ in the global queue (sorted by arrival time) \;  \BlankLine
 The local queues of the GPUs \;  \BlankLine
 \Indm
}
 \BlankLine
Foreach $GPU_i$ in $idle\_GPUs$ \{ \;  \BlankLine
        \Indp 
        \tcc{Prioritize the requests in the local queue already scheduled to $GPU_i$ }
        If $GPU_i$ 's local queue is not empty \{ \;  \BlankLine
                \Indp 
                Dispatch the request at the head of the local queue to $GPU_i$ \;  \BlankLine
                \textbf{Continue} \;  \BlankLine
                \Indm
        \} \;  \BlankLine
        
        \tcc{Look for a request in the global queue that has its model cached on $GPU_i$}
        Foreach $Req_s$ in $requests$ \{ \;  \BlankLine
             \Indp 
             If $Req_s$’s model is cached in $GPU_i$ \{ \;  \BlankLine
                    \Indp 
                      Dispatch $Req_s$ to $GPU_i$ \;  \BlankLine
                      \textbf{Break} \;  \BlankLine
                    \Indm
             \} \;   \BlankLine
            \tcc{Enforce the out-of-order dispatch limit (the limit is set to 0 if out-of-order dispatch is disabled}
             If the number of visits of $Req_s$ is higher than the specified limit \{ \;   \BlankLine
                    \Indp 
                      $flag$ = \textbf{LocalityLoadBalance}($GPU_i$, $idle\_GPUs$, $busy\_GPUs$, $Req_s$) \;  \BlankLine
                      If $flag$ is True \textbf{Break} Else \textbf{Continue} \;  \BlankLine
                    \Indm
             \} Else \{ \;  \BlankLine
                    \Indp 
                      Increment the number of visits of $Req_s$ \;  \BlankLine
                    \Indm
             \} \; 
            \Indm
        \} Else \{ \;  \BlankLine
            \Indp 
            \tcc{No request in the global queue has its model cached on $GPU_i$}
            Foreach $Req_s$ in $requests$ \{ \;  \BlankLine
                    \Indp 
                     $flag$ = \textbf{LocalityLoadBalance}($GPU_i$, $idle\_GPUs$, $busy\_GPUs$, $Req_s$) \;  \BlankLine
                     If $flag$ is True \textbf{Break} Else \textbf{Continue} \;  \BlankLine
                    \Indm
            \} \;  \BlankLine
            \Indm
        \} \;  \BlankLine
        \Indm
\} \;  \BlankLine  
\caption{\textbf{LALB:} Locality-Aware Load-Balancing}
\label{alg:lalbo3}
\end{algorithm}

\begin{algorithm}
\DontPrintSemicolon
\BlankLine
\KwInput{ 
 \;
 \Indp 
 The selected idle GPU $GPU_i$ \; \BlankLine
 The list of idle GPUs $idle\_GPUs$ \; \BlankLine
 The list of busy GPUs $busy\_GPUs$ \; \BlankLine
 The selected request $Req_s$ \; \BlankLine
 \Indm
}
\KwOutput{ 
 \;
 \Indp 
 A Boolean value indicating whether $Req_s$ is dispatched to $GPU_i$ (True) or not (False) \; \BlankLine
 \Indm
}
\tcc{Allow cache miss if $Req_s$'s model is not cached on any idle or busy GPU}
If $Req_s$'s model is not cached on any other GPU \{ \; \BlankLine     
        \Indp 
        Dispatch $Req_s$ to $GPU_i$ \; \BlankLine
        \textbf{Return True} \;  \BlankLine
        \Indm
\} Else if $Req_s$'s model is cached on another idle GPU $GPU_j$ \{ \; \BlankLine
        \Indp 
        Dispatch $Req_s$ to $GPU_j$ \;\BlankLine 
        \textbf{Return False} \;\BlankLine 
        \Indm
\}  Else \{ \;\BlankLine 
        \Indp 
        \tcc{$Req_s$'s model is cached on one of the busy GPUs}
        Foreach $GPU_j$ in $busy\_GPUs$ \{ \; \BlankLine
            \Indp
            If $GPU_j$ has $Req_s$'s model cached \{ \; \BlankLine
                \Indp 
                Estimate the finish time of the requests on $GPU_j$ \;\BlankLine
                If the finish time is less than $Req_s$'s model loading time \{ \;\BlankLine 
                  \Indp 
                  Move $Req_s$ to $GPU_j$'s local queue  \;\BlankLine
                  \textbf{Return False} \;\BlankLine
                  \Indm
                \} \;\BlankLine
                \Indm
            \} \;\BlankLine
            \Indm
        \} Else \{ \;\BlankLine
                \Indp 
                \tcc{Allow cache miss}
                Dispatch $Req_s$ to $GPU_i$ \;\BlankLine 
                \textbf{Return True} \;\BlankLine
                \Indm
        \} \;\BlankLine
        \Indm
\} \;\BlankLine
\caption{ Function  \textbf{LocalityLoadBalance}}
\label{alg:localfunc}
\end{algorithm}

\subsection{Out of Order Dispatch}

The LALB scheduler also adopts an out-of-order (O3) dispatch policy to further promote the reuse of models on the GPUs. It prioritizes a waiting request that can be cache hit on an idle GPU by dispatching it to the idle GPU ahead of the requests queued in front of it in the global queue (Algorithm \ref{alg:lalbo3} Lines 6-10).
By default, the LALB scheduler dispatches the requests in the global queue in the order of their arrivals.
In-order dispatch ensures fairness among the requests, but it may lead to suboptimal performance because a dispatched request may evict a model that is needed soon by a following request.
The negative impact of such occurrences can be severe especially when the workload's working set (the total number of unique models) is large and can lead to thrashing behaviors.
Out-of-order dispatch sacrifices a certain level of fairness in exchange for much improved performance.
To avoid starvation, it sets a specified limit (by default 25) to prevent waiting requests from being starved (Algorithm \ref{alg:lalbo3} Lines 11-14).
If the number of times a request in the queue has been skipped meets this limit, the scheduler will dispatch the request immediately, regardless of whether it causes a cache hit or miss.

\section{Evaluation}
\label{sec:evaluation}

\subsection{Methodology}

We evaluate the performance of our proposed GPU-enabled FaaS system using real-world ML inference workloads. The prototype is implemented upon OpenFaaS. 
We consider three metrics for evaluating the proposed GPU-enabled FaaS: average latency, cache miss ratio, and GPU (SM) utilization.
To show the improvement made by our proposed cache-locality-aware scheduler, we compare its performance to the default load-balancing scheduler, which simply dispatches the request at the head of the global queue whenever a GPU becomes idle.



\subsubsection{Workloads}

Table \ref{tab:model_table} shows the 22 popular CNN models considered in our workload.
The table lists the models' actual size and the occupation size in GPU memory when the model inference runs with the fixed batch size of 32. 
The Cache Manager uses this peak memory occupation size for the cache replacement decision, as the GPU would result in OOM if it exceeds the available memory. 

We use the Microsoft Azure function trace \cite{shahrad2020serverless} to evaluate the performance of the schedulers. 
We choose this trace because it represents the workload of FaaS functions provisioned by a real-world major cloud provider, Microsoft Azure. 
The trace contains 14 files representing 14 days of function invocations. 
Each file provides a column representing each minute, a row representing each unique function, and a value indicating the total invocations of the unique function per minute.  
We extract the first 6 minutes of the trace and normalize the number of requests for each minute to 325 requests to match the size of our much smaller testbed of 12 GPUs that we use for our experiment.

The total number of unique functions (working set) in the Azure trace is 46,413, which is too large for our testbed to handle.   
At the same time, the trace represents a workload with a highly skewed working set: the top 15 popular functions together represent 56\% of the total invocations per minute, whereas the functions below the top 15 each represent less than 0.01\% of the total invocations per minute. 
Therefore, we consider only the most frequently used functions as the working set in our workload.
We use three different working set sizes: 15, 25, and 35.
A larger working set introduces more unique functions while maintaining the maximum number of requests per minute at 325 requests. 
We map each unique function in the trace to a unique model in Table~\ref{tab:model_table} and ensure models with different sizes are distributed evenly in the workload.
Within each minute of the workload, we randomly distribute the invocations of different functions while maintaining the normalized total invocations per minute.

\begin{table}[t]
\centering
\resizebox{\columnwidth}{!}{%
\begin{tabular}[t]{p{0.15\textwidth} p{0.15\textwidth} p{0.10\textwidth} p{0.10\textwidth} }
\toprule
Model & Size (MB) & Loading time (s) &  Inference time (s) \\
\midrule
squeezenet1.1	       & 1269	& 2.41	 & 1.28 \\
resnet18	           & 1313	& 2.52	 & 1.25 \\
resnet34	           & 1357	& 2.60	 & 1.25 \\
squeezenet1.0	       & 1435	& 2.32	 & 1.33 \\
alexnet	               & 1437	& 2.81	 & 1.25 \\
resnext50.32x4d	       & 1555	& 2.64	 & 1.29 \\
densenet121	           & 1601	& 2.49	 & 1.28 \\
densenet169	           & 1631	& 2.56	 & 1.30 \\
densenet201	           & 1665	& 2.67	 & 1.40 \\
resnet50	           & 1701	& 2.67	 & 1.28 \\
resnet101	           & 1757	& 2.95	 & 1.30 \\
resnet152	           & 1827	& 3.10	 & 1.31 \\
densenet161	           & 1919	& 2.75	 & 1.32 \\
inception.v3	       & 2157	& 4.42	 & 1.63 \\
resnext101.32x8d       & 2191	& 3.51	 & 1.33 \\
vgg11	               & 2903	& 3.94	 & 1.29 \\
wide\_resnet50\_2	   & 3611	& 3.16   & 1.31 \\
wide\_resnet101\_2     & 3831	& 3.91	 & 1.32 \\
vgg13	               & 3887	& 3.98	 & 1.30 \\
vgg16	               & 3907	& 4.04	 & 1.27 \\
vgg16.bn               & 3907	& 4.03	 & 1.26 \\
vgg19                  & 3947	& 4.07   & 1.33 \\
\bottomrule
\end{tabular}
}
\caption{Occupation size in GPU, loading time, and inference latency (for a batch size of 32) of the models used in the evaluation}
\vspace{-12pt}
\label{tab:model_table}
\end{table}

\subsubsection{Dataset}

For the input images used for inference, we provide a small group of 150 image files which comprise standard datasets such as CIFAR10 \cite{krizhevsky2009learning}, Modified National Institute of Standards, and Technology (MNIST) \cite{mnist}, and Hymenoptera \cite{pytorch}. The MNIST dataset provides 28x28 grayscale images that splits into 60,000 training and 10,000 validation images. 
CIFAR-10 includes 32x32 RGB images that have 50,000 training images and 10,000 validation images.
Hymenoptera provides RGB images ranging from 50KB to 2MB in size that must be compressed before being used in model inference. The dataset consists of 245 training images and 153 testing images.

\subsubsection{Testbed.}

We conducted all the experiments on three GPU servers, each equipped with four GeForce RTX 2080 GPUs, and deployed GPU Managers as Nvidia Docker containers with access to GPU resources.
A separate server is used to run Docker containers for Scheduler, Datastore, and Cache Manager, and the components required for the OpenFaaS platform.
Every server has dual Intel Xeon 8-core processors and 128GB DRAM, and they are interconnected by a 100Gb/s InfiniBand network. Data is stored on an NFS-based shared file system backed by a 6TB SATA HDD.

\begin{figure*}[t]
     \centering
     \begin{subfigure}[b]{0.3\textwidth}
         \centering
         \includegraphics[width=0.9\textwidth]{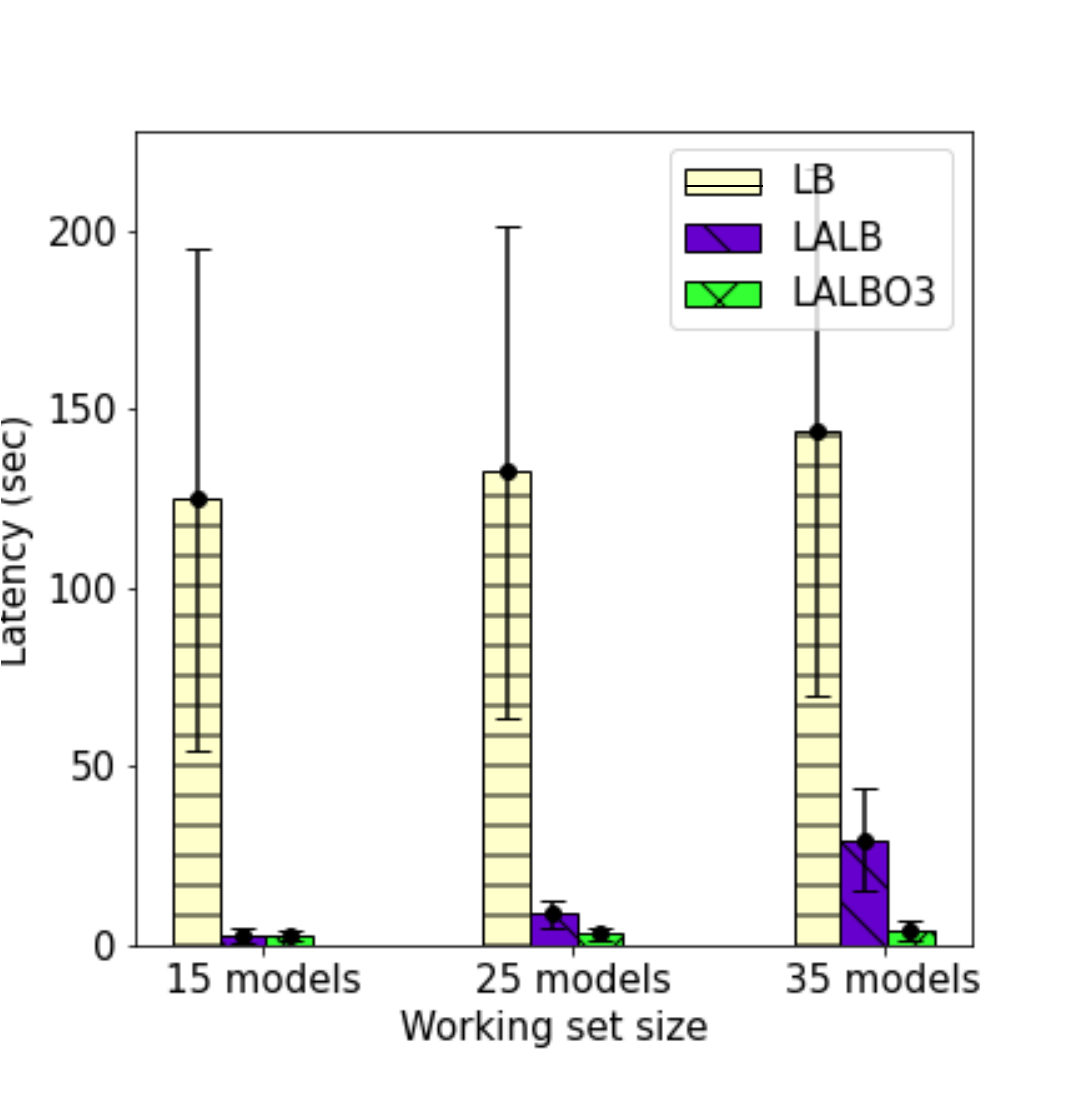}
         \vspace{-9pt}
         \caption{Average Function Latency}
         \label{fig:totalAverageLatency}
     \end{subfigure}
     \hfill
     \begin{subfigure}[b]{0.3\textwidth}
         \centering
         \includegraphics[width=0.9\textwidth]{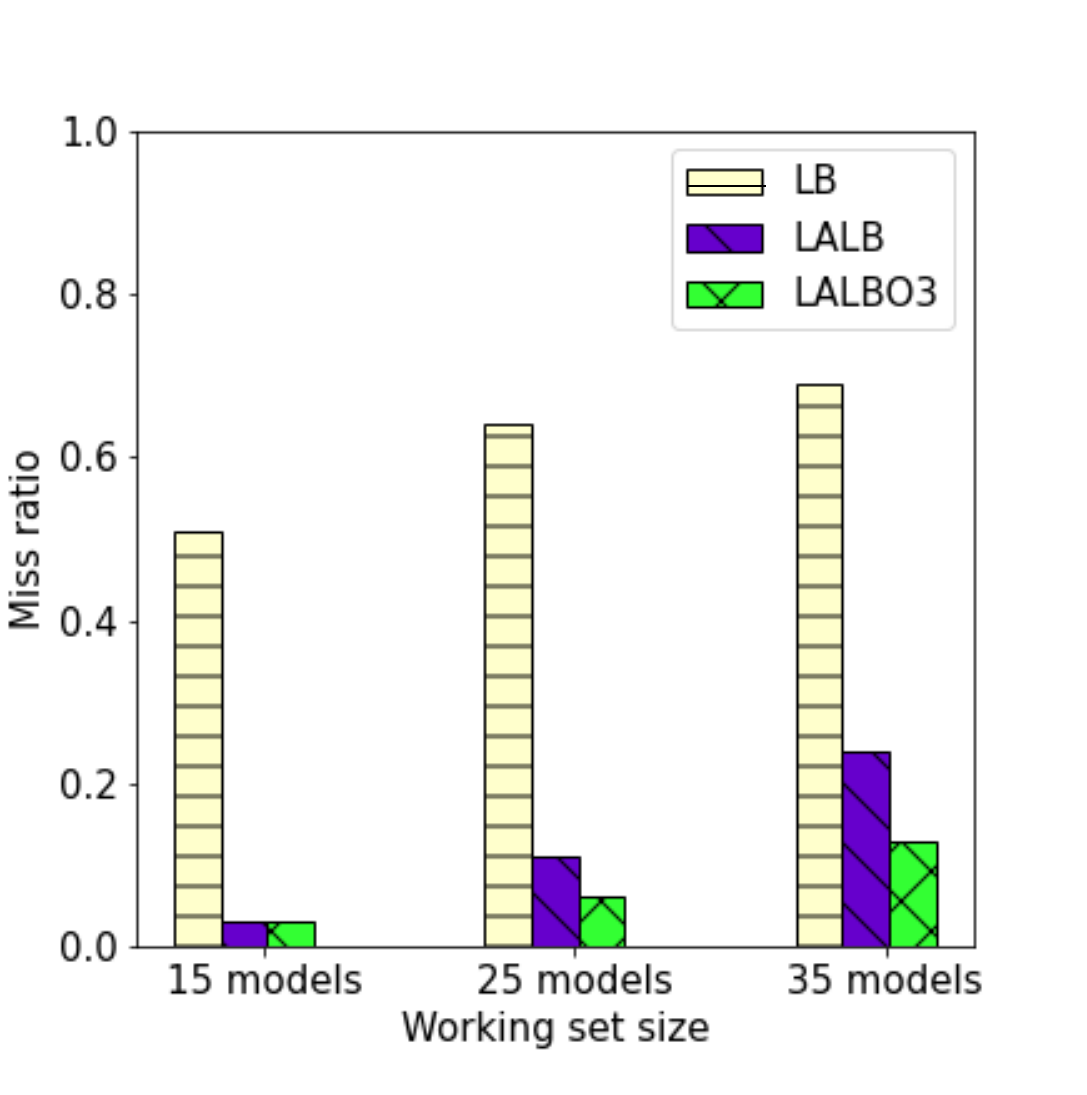}
         \vspace{-9pt}
         \caption{Cache Miss Ratio}
         \label{fig:cacheMissRatio}
     \end{subfigure}
     \hfill
     \begin{subfigure}[b]{0.3\textwidth}
         \centering
         \includegraphics[width=0.9\textwidth]{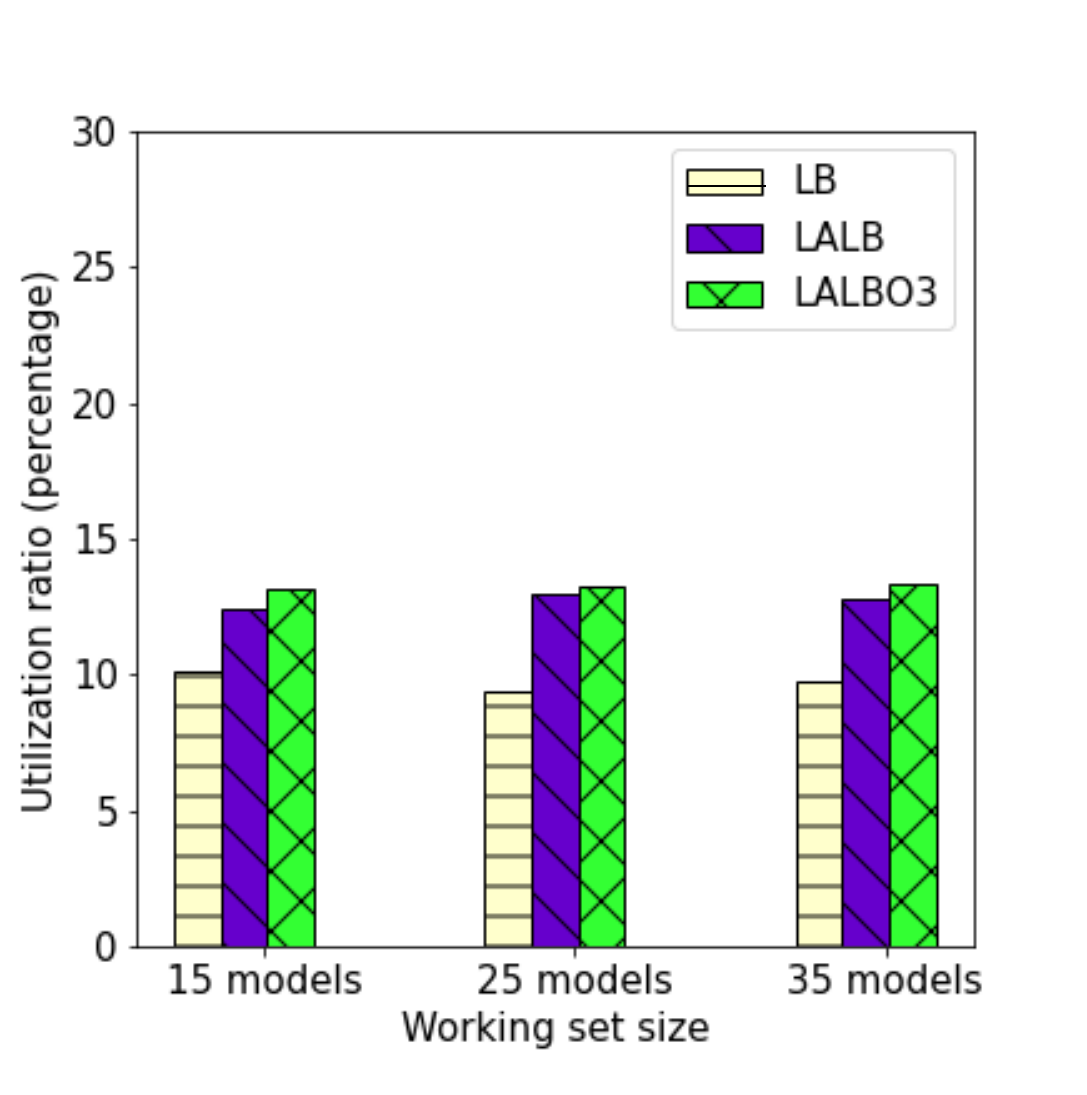}
         \vspace{-9pt}
         \caption{GPU (SM) Utilization}
         \label{fig:gpuSmUtlization}
     \end{subfigure}
        \caption{Comparative analysis of average latency, cache miss ratio, GPU (SM) utilization between the proposed schedulers and the default load-balancing scheduler}
        \label{fig:EvalFig}
        \vspace{-12pt}
\end{figure*}

\subsection{Latency and Cache Miss Ratio}

Figure \ref{fig:totalAverageLatency} shows the total function latency of the three schedulers. 
The results show that the Locality-Aware-Load-Balancing (LALB) scheduler reduces the average latency by 97.74\% and 93.33\% compared to the default Load-Balancing (LB) scheduler with working set sizes of 15 and 25 respectively.
However, the average latency and the cache miss ratio of the LALB scheduler degrade as the working set size increases to 35.
The result indicates that the cache miss ratio reduces by 94.11\% with the working set size of 15 but reduces by 65.21\% with the working set size of 35, as shown in Figure \ref{fig:cacheMissRatio}. 
The degrading performance is because improving locality becomes challenging when the working set size becomes more extensive.

Applying the Out-of-Order (O3) dispatch to the LALB scheduler further improves the performance with the working set size of 25 and 35. 
The O3 dispatch promotes cache hits by allowing requests in the global queue to be dispatched out of order if they can generate cache hits. 
As the working set size increases and overwhelms the limited GPU memory, reducing the cache miss ratio becomes essential. 
The LALB scheduler reduces the cache miss ratio of LB by 65.21\%, and the LALBO3 scheduler reduces the cache miss ratio of LB by 81.15\% with the working set size of 35.  

\subsection{Utilization}

Figure \ref{fig:gpuSmUtlization} shows the average SM utilization of all the GPUs when running the workloads using the different schedulers. 
The SM utilization of each scheduler remains consistent across all three working sets, as the maximum number of requests per minute is always kept at 325. 
Reaching the SM utilization of 100\% is impossible as the GPUs accommodate multiple inference models and cannot risk exceeding memory by allocating a too large batch size.

The LALBO3 scheduler has the highest SM utilization due to the lowest cache miss ratio.
The SM utilization negatively correlates with the cache miss ratio because a GPU cannot use the SM to perform inference until the model is uploaded to the GPU memory. 
When there is a cache-miss, the SM utilization remains zero until the victim model becomes evicted and the new model is uploaded to the GPU.
As a result, the LALBO3 scheduler shows the highest SM utilization as it has the lowest cache miss ratio.

\begin{figure}
\begin{minipage}[t]{0.5\columnwidth}
  \includegraphics[width=\linewidth]{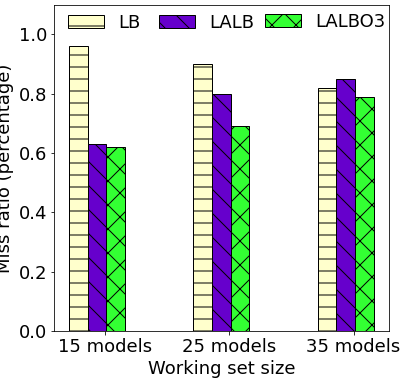}
  \caption{False Miss Ratio}
  \label{fig:falseMissRatio}
\end{minipage}\hfill 
\begin{minipage}[t]{0.5\columnwidth}
  \includegraphics[width=\linewidth]{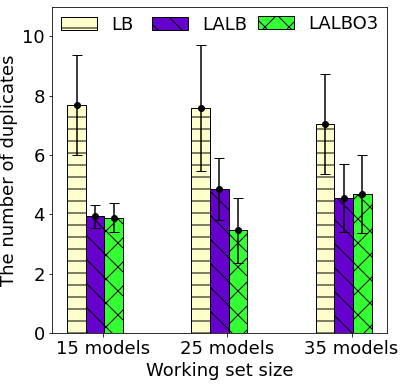}
  \caption{Average number duplicates of the top one model}
  \label{fig:duplicates}
\end{minipage}
\vspace{-9pt}
\end{figure}

\subsection{Efficiency}

This section explains the efficiency of the schedulers using the false miss ratio and the number of duplicated hot models as the metrics. 
Having duplicates of hot models in the system allow the inference requests for these models to generate hits, but having too many duplicates can also pollute the cache and reduce the overall hit ratio.
The ideal scheduler should maintain a minimal number of duplicated models on the GPUs without degrading the cache hit ratio.
The average number of duplicates is collected by tracking the total number of GPUs that has the most popular model cached at the same time during the experiment.
A false miss is a cache miss scenario in the scheduling decision where the request is forwarded to a GPU as a cache miss even though the requested model is cached on another GPU. 
The false miss ratio correlates with the number of duplicated models because the false miss decisions force the GPUs without the requested models to store the models that are already held on the other GPUs.

Figure \ref{fig:falseMissRatio} shows that both LALB and LALBO3 schedulers reduce the false miss ratio with working set sizes of 15 and 25.
The default LB schedule naturally has the worst false miss ratio (as high as nearly 96\%) since it does not consider cache locality when scheduling requests.
Compared to LB, for the working set size of 15 LALB and LALBO3 reduce the false miss ratio by 34.38\% and 35.41\%, respectively. 
The available GPU memory can find the optimal number of duplicated cache items for the small working set to promote locality without the O3 dispatch. 
As the working set size increases to 35, only the LALBO3 scheduler can still reduce the false miss ratio of the LB scheduler by 3.65\%, as LALBO3 has the O3 dispatch to exploit locality further by promoting the waiting requests to the cached GPUs. 

Figure \ref{fig:duplicates} shows the average number of duplicates for the most popular model. 
As the GPU-enabled FaaS uses 12 GPUs, the highest number of duplicates of the same model cannot exceed 12. 
As the LB scheduler does not consider locality, it is subject to the situation where the duplicated cache items continuously evict each other.
%
The LALB scheduler improves locality by judiciously selecting GPUs with cached models to process requests, and the increased cache hits reduce the number of duplicated cache items in the system.
The LALB scheduler reduces the average number of duplicates of LB  by 48.96\% with the working set size of 15 while reducing the cache miss ratio by 94.11\% (shown in Figure~\ref{fig:cacheMissRatio}). 
Increasing the working set size degrades the ability of the LALB scheduler to maintain the optimal number of duplicates, and it reduces the duplicates of LB by 35.32\% with a working set size of 35. 
Larger working sets make it difficult for LALB to maintain cache locality, given the limited GPU memory capacity. 

With the working set size of 15, the LALBO3 scheduler does not significantly reduce the number of duplicates compared to LALB, as it reduces the average number of duplicates of the LB scheduler by 49.48\%.
It shows negligible performance improvement for the O3 dispatch because the available GPU memory is enough to cover most of the working set. 
The LALBO3 scheduler performs better than the LALB scheduler with a working set size of 35 by reducing the average number of duplicates of the LB scheduler by 33.47\%.
The results indicate that by applying the O3 dispatch, locality performance can be further improved to reduce the average number of duplicates.

\begin{figure}
\centering
\includegraphics[width=\linewidth]{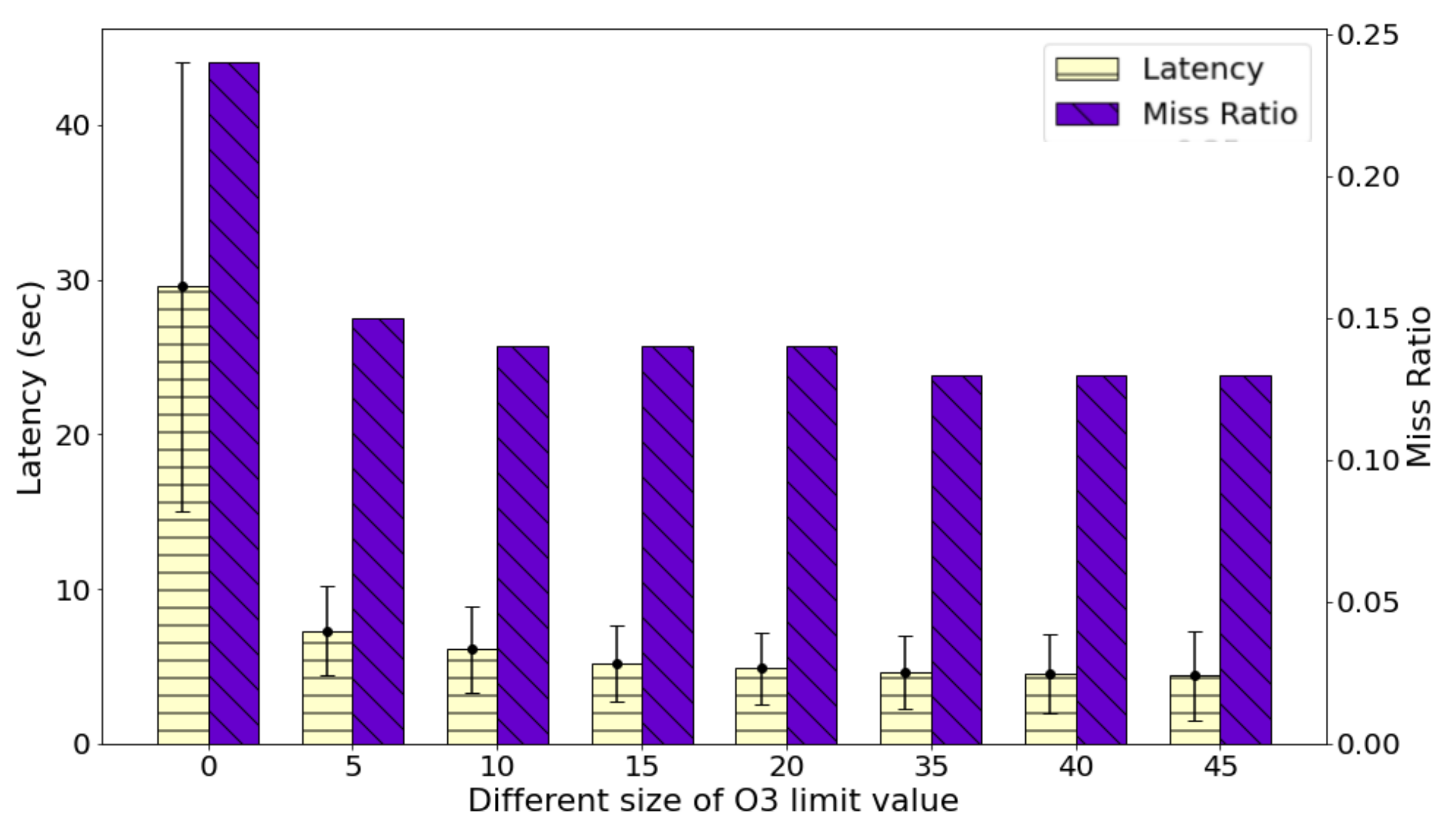}
\vspace{-12pt}
\caption{Average function latency and cache miss ratio of different limit value used by the out-of-order (O3) dispatch}
\vspace{-18pt}
\label{fig:oolimit}
\qquad
\end{figure}


\subsection{O3 Sensitivity Test}

Figure \ref{fig:oolimit} focuses on the sensitive study of the limit value used by the O3 dispatch in the LALBO3 scheduler, which determines how many times a request in the queue can be skipped in order to promote a later request that has the model cached in the idle GPUs. 
We experiment with the workload with the working set size of 35 and change the specified limit of the O3 dispatch from zero to 45 (x-axis).
With the limit set to zero, LALBO3 reduces to LALB; a higher limit value, on the other hand, can potentially cause more unfairness in request dispatches.
The average function latency (left y-axis) and the cache miss ratio (right y-axis) are used to evaluate the performance changes created by the different limit values. 

The results show that both the latency and cache miss ratio reduce as we increase the specified limit value of O3. 
%
The O3 limit of 45 reduces the average latency and cache miss ratio of the O3 limit value of 0 by 85.1\% and 45.83\%, respectively.
The larger O3 value increases the performance because it prioritizes cache hits over the arrival order while dispatching requests. 
Interestingly, the O3 limit value of 45 also reduces, instead of increasing, the variance of the average latency of the limit value of 0 by 95.93\%.
On one hand, a larger limit value causes some requests to wait unfairly longer; on the other hand, a lower limit value causes more cache misses.
The results show that the latter dominates the latency variance, and as such the larger limit value significantly reduces the latency variance compared to the limit value of zero, i.e., LALB.

\section{Discussions}
\label{sec:discussions}

\textbf{Overhead and Scalability.} 
The GPU Managers are distributed, one per GPU node, and can therefore scale with the number of nodes. On each node, the GPU Manager can scale with the number of models by using multiple GPU processes to handle the concurrent models that the local GPUs need to serve, one per model.

The Cache Manager is global and its scalability is ensured by that each GPU’s memory is managed separately (using a separate LRU list). The cache manager can use multiple threads to handle concurrent requests from multiple GPU Managers in parallel.

The Scheduler is global, and it runs the locality-aware load-balancing (LALB) algorithm to make scheduling decisions for model inference functions. 
When a GPU becomes idle, the Scheduler searches the global queue for a request that already has its model cached in the GPU. To reduce this search overhead, the Scheduler maintains an auxiliary data structure that links the queued requests to their corresponding models---the requests linked to the same model are still sorted by their arriving order. By using this data structure, the complexity of this search is bounded by the number of models cached on the GPU.

When a request needs to be scheduled, the Scheduler checks all the GPUs that have this request's model cached---if it finds an idle GPU, the request is immediately dispatched there; if it finds a busy GPU that can serve this request sooner that an idle one, the request is moved to this busy GPU's local queue. To facilitate this search, the Cache Manager maintains the lists of GPUs where each model is cached, and shares this information with the Scheduler through the Datastore (Etcd). Therefore, the complexity of this search is bounded by the number of GPUs that have this model cached.

\textbf{Multi-tenancy and Security:} 
Our solution leverages the underlying FaaS framework to support multi-tenancy. For example, OpenFaaS Pro has support for multiple-namespaces~\cite{openfaas-pro}, which in combination with its security features, can provide logical segregation of groups of functions belonging to different tenants.

As our solution enables model inference functions from different tenants to share the GPUs, we also need to provide isolation on the GPU resources.
A bad actor may attempt to overload the system by submitting many inference requests; our system can address this and provide isolation by limiting the number of GPU processes that each tenant can use. A bad actor may also attempt to game the system by designing a high-locality workload and monopolizing the GPU time and memory; our system can address this and provide isolation by limiting the GPU time share and memory space share that a tenant can use.

\textbf{Heterogeneity of GPUs:} Our solutions can inherently support the use of heterogeneous GPUs for model inference functions. It just needs to run the same profiling procedure described in Section~\ref{sec:lalb} on each unique type of GPUs in the system, and use the profiled model loading and inference times in the proposed scheduling algorithm.

\textbf{Cache Replacement Policy: }
Our current implementation uses LRU as the policy to manage the replacement of cached models in each GPU.
Our system’s design can easily support other cache replacement policies (by replacing the LRU lists with other types of sorted lists). But regardless of what policy is used, our proposed locality-aware scheduling can always improve its performance by promoting the reuse of cached models and boosting cache hits while maintaining load balance.

\textbf{Usability.} 
Our solution is built upon the commonly used FaaS and machine learning frameworks and users do not need to make any changes to their code. The modifications that our solution makes to the FaaS framework and the users' Dockerfiles are small and entirely transparent to the users.
At the same time, our proposed techniques for cache-locality-aware load balancing of model inference tasks and our implementation of the Scheduler, GPU Manager, and Cache Manager can be adopted by other model serving systems (e.g., TensorFlow Serving~\cite{tensorflow-serving}, TorchServe~\cite{torchserv}) to improve their performance of GPU-based model inference.

\section{Conclusions}
\label{sec:conclusions}

The demand for GPU-enabled FaaS is growing as the use cases of ML inference tasks can significantly benefit from GPU acceleration and function-based deployment and execution.
Our solution focuses on improving the FaaS functions running ML inference tasks such as CNN that can significantly benefit from GPU acceleration. 
However, the existing FaaS frameworks provide limited support for FaaS functions to share GPU resources, and the short-lived nature of FaaS functions makes it difficult to overcome the cost of data transfer for model inference on GPUs.

Our proposed solution can be applied to different FaaS frameworks, requiring only a few complementary components to introduce GPU scheduling and cache management. 
Our GPU-enabled FaaS provides global management of GPU memory and treats the uploaded inference models in GPUs as cache items to reduce the data transfer overhead. 
Our cache-locality-aware scheduler considers both GPU cache locality and load balance to improve the performance of model inference functions.


We have used real-world FaaS trace and ML models widely used in production to evaluate the performance of our GPU-enabled FaaS solution. The results show that it can substantially improve model inference performance.
For example, the LALB scheduler reduces the baseline (LB) scheduler's average latency and cache miss ratio by 79.43\% and 65.21\%, respectively, for a workload of working set size of 35.
Additionally, the out-of-order (O3) dispatch can work with the LALB scheduler to further improve the locality performance and reduce the LB scheduler's average latency and cache miss ratio by 96.93\% and 81.16\%, respectively.


\bibliographystyle{plain}

\begin{thebibliography}{10}

\bibitem{tensorflow}
Mart{\'\i}n Abadi, Ashish Agarwal, Paul Barham, Eugene Brevdo, Zhifeng Chen,
  Craig Citro, Greg~S Corrado, Andy Davis, Jeffrey Dean, Matthieu Devin, et~al.
\newblock Tensorflow: Large-scale machine learning on heterogeneous distributed
  systems.
\newblock {\em arXiv preprint arXiv:1603.04467}, 2016.

\bibitem{aws-vgpu}
Virtual {GPU} device plugin for inference workloads in {Kubernetes}.
\newblock
  \url{https://aws.amazon.com/blogs/opensource/virtual-gpu-device-plugin-for-inference-workload-in-kubernetes/}.

\bibitem{awslambda}
Poornima Chand.
\newblock Machine learning inference at scale using {AWS} serverless.
\newblock {\em Amazon blog}, 11 2021.

\bibitem{funcx}
Ryan Chard, Yadu~N. Babuji, Zhuozhao Li, Tyler~J. Skluzacek, Anna Woodard, Ben
  Blaiszik, Ian~T. Foster, and Kyle Chard.
\newblock {funcX}: {A} federated function serving fabric for science.
\newblock {\em CoRR}, abs/2005.04215, 2020.

\bibitem{8814494}
Abdul Dakkak, Cheng Li, Simon Garcia~de Gonzalo, Jinjun Xiong, and Wen-mei Hwu.
\newblock Trims: Transparent and isolated model sharing for low latency deep
  learning inference in function-as-a-service.
\newblock In {\em 2019 IEEE 12th International Conference on Cloud Computing
  (CLOUD)}, pages 372--382, 2019.

\bibitem{docker-swarm}
{Swarm mode overview}.
\newblock \url{https://docs.docker.com/engine/swarm/}.

\bibitem{5547126}
José Duato, Antonio~J. Peña, Federico Silla, Rafael Mayo, and Enrique~S.
  Quintana-Ortí.
\newblock {rCUDA}: Reducing the number of {GPU}-based accelerators in high
  performance clusters.
\newblock In {\em 2010 International Conference on High Performance Computing
  Simulation}, pages 224--231, 2010.

\bibitem{etcdsite}
Etcd: A distributed, reliable key-value store for the most critical data of a
  distributed system.
\newblock 2021.

\bibitem{azfunction}
Anirudh Garg.
\newblock Why use {Azure} functions for {ML} inference?
\newblock {\em Microsoft blog}, 05 2020.

\bibitem{jin2020faas}
Runyu Jin, Qirui Yang, and Ming Zhao.
\newblock Is {FaaS} suitable for edge computing.
\newblock {\em USENIX Association, June}, 2020.

\bibitem{8374513}
Jaewook Kim, Tae~Joon Jun, Daeyoun Kang, Dohyeun Kim, and Daeyoung Kim.
\newblock {GPU} enabled serverless computing framework.
\newblock In {\em 2018 26th Euromicro International Conference on Parallel,
  Distributed and Network-based Processing (PDP)}, pages 533--540, 2018.

\bibitem{knative}
{Knative}.
\newblock \url{https://github.com/knative/docs}.

\bibitem{koliousis2019crossbow}
Alexandros Koliousis, Pijika Watcharapichat, Matthias Weidlich, Luo Mai, Paolo
  Costa, and Peter Pietzuch.
\newblock Crossbow: Scaling deep learning with small batch sizes on multi-{GPU}
  servers.
\newblock {\em arXiv preprint arXiv:1901.02244}, 2019.

\bibitem{krizhevsky2009learning}
Alex Krizhevsky, Geoffrey Hinton, et~al.
\newblock Learning multiple layers of features from tiny images.
\newblock 2009.

\bibitem{kubernetes}
{Kubernetes}.
\newblock \url{https://kubernetes.io/}.

\bibitem{mnist}
Yann LeCun.
\newblock The {MNIST} database of handwritten digits.
\newblock \url{http://yann. lecun. com/exdb/mnist/}.

\bibitem{mps}
Multi-process service ({MPS}).
\newblock \url{https://docs.nvidia.com/deploy/mps/index.html}.

\bibitem{10.1016/j.jpdc.2020.01.004}
Diana~M. Naranjo, Sebasti\'{a}n Risco, Carlos de~Alfonso, Alfonso P\'{e}rez,
  Ignacio Blanquer, and Germ\'{a}n Molt\'{o}.
\newblock Accelerated serverless computing based on {GPU} virtualization.
\newblock {\em J. Parallel Distrib. Comput.}, 139(C):32–42, May 2020.

\bibitem{nvidiadocker}
Nvidia docker container toolkit.
\newblock
  \url{https://docs.nvidia.com/datacenter/cloud-native/container-toolkit/install-guide.html}.

\bibitem{openfaas-pro}
{OpenFaaS Pro} namespaces support.
\newblock \url{https://docs.openfaas.com/reference/namespaces/}.

\bibitem{pytorch}
{PyTorch}.
\newblock \url{https://pytorch.org/}.

\bibitem{satzke2020efficient}
Klaus Satzke, Istemi~Ekin Akkus, Ruichuan Chen, Ivica Rimac, Manuel Stein,
  Andre Beck, Paarijaat Aditya, Manohar Vanga, and Volker Hilt.
\newblock Efficient {GPU} sharing for serverless workflows.
\newblock In {\em Proceedings of the 1st Workshop on High Performance
  Serverless Computing}, pages 17--24, 2020.

\bibitem{shahrad2020serverless}
Mohammad Shahrad, Rodrigo Fonseca, {\'I}{\~n}igo Goiri, Gohar Chaudhry, Paul
  Batum, Jason Cooke, Eduardo Laureano, Colby Tresness, Mark Russinovich, and
  Ricardo Bianchini.
\newblock Serverless in the wild: Characterizing and optimizing the serverless
  workload at a large cloud provider.
\newblock In {\em 2020 USENIX Annual Technical Conference (USENIX ATC 20)},
  pages 205--218, 2020.

\bibitem{tensorflow-serving}
Tensorflow serving.
\newblock \url{https://www.tensorflow.org/tfx/guide/serving}.

\bibitem{torchserv}
Torchserv.
\newblock \url{https://pytorch.org/serve/}.

\bibitem{triton}
Nvidia {Triton} inference server.
\newblock \url{https://developer.nvidia.com/nvidia-triton-inference-server}.

\bibitem{wolf2020transformers}
Thomas Wolf, Lysandre Debut, Victor Sanh, Julien Chaumond, Clement Delangue,
  Anthony Moi, Pierric Cistac, Tim Rault, R{\'e}mi Louf, Morgan Funtowicz,
  et~al.
\newblock Transformers: State-of-the-art natural language processing.
\newblock In {\em Proceedings of the 2020 conference on empirical methods in
  natural language processing: system demonstrations}, pages 38--45, 2020.

\bibitem{xiao2020antman}
Wencong Xiao, Shiru Ren, Yong Li, Yang Zhang, Pengyang Hou, Zhi Li, Yihui Feng,
  Wei Lin, and Yangqing Jia.
\newblock {AntMan}: Dynamic scaling on {GPU} clusters for deep learning.
\newblock In {\em 14th USENIX Symposium on Operating Systems Design and
  Implementation (OSDI 20)}, pages 533--548, 2020.

\bibitem{yeh2020kubeshare}
Ting-An Yeh, Hung-Hsin Chen, and Jerry Chou.
\newblock {KubeShare}: A framework to manage {GPU}s as first-class and shared
  resources in container cloud.
\newblock In {\em Proceedings of the 29th International Symposium on
  High-Performance Parallel and Distributed Computing (HPDC)}, pages 173--184,
  2020.
\end{thebibliography}

\end{document}